# Memristor Load Current Mirror Circuit

Olga Krestinskaya, Irina Fedorova, and Alex Pappachen James
School of Engineering
Nazarbayev University
Astana, Republic of Kazakhstan

*Abstract*— Simple current mirrors with semiconductor resistive loads suffer from large on-chip area, leakage currents and thermal effects. In this paper, we report the feasibility of using memristive loads as a replacement of semiconductor resistors in simplistic current mirror configuration. We report power, area and total harmonic distribution, and report the corner conditions on resistance tolerances.

## I. INTRODUCTION

Current mirrors are used to mirror the reference current across different part of a given circuit and replaces the need to have multiple current sources. They are useful in analog circuits when the reduction of current variation is a primary requirement for using current mirrors [1]. They find applications in differential amplifiers, operational amplifiers, adder circuits, multipliers etc. [2]. Although, the wide range of applications make current mirror quite important, the simple current mirror has several practical limitations.

One of the major limitation in a simple current mirror with resistive loads is the practical realization of resistors is the cost of large area on chip, reverse leakage currents and large variability in resistors values. To counteract these issues, we propose to replace resistors with resistance emulating memristive devices. The low leakage currents, very low on-chip area and programmability of resistance value make memristors an excellent replacement for semiconductor resistors [3]. In the next section we quickly review the memristor model and verify the hysteresis behavior of memristor for different frequencies. After this, section 3 shortly describes basic current mirror circuits with the semiconductor resistor and memristor models that will be further under examination. Section 4 and 5 describes the simulation results and the main parameters of the circuits, namely total harmonic distortion, area and power dissipation. Section 6 provides conclusions.

## II. MEMRISTOR MODEL

The memristor is considered as the fourth canonical circuit element, where three others are resistor, capacitor and inductor. Its static relationship is defined by (1), where $\varphi$ is the flux and $q$ is the charge.

$$M = \frac{d\varphi}{dq} \qquad (1)$$

The behavior of the memristor is similar to the series combination of the two variable resistors. Moreover, memristor could be described by dividing it into two regions with the moving boundary wall at position $w$. Consider the memristor of height of $L$, height w corresponds to the region of memristor with low resistance and rest of the height L-W represents the high resistance region [4], then the total effective memristor resistance could be written as (2):

$$M(t) = \frac{W}{L}R_{on} + \left(1 - \frac{W}{L}\right)R_{off} \qquad (2)$$

One of the important features describing the behavior of memristor is shown by the 'pinched hysteresis loop' (Fig. 1). It is important to note that the curve passes through zero. We performed the simulation of memristor characteristics for the range of frequencies to verify memristor behavior. As the frequency increases the hysteresis loop narrows down because the difference between the $R_{on}$ and $R_{off}$ reduces. For the high frequencies $R_{on}$ and $R_{off}$ get the same value, which could be shown as a straight line on the graph.

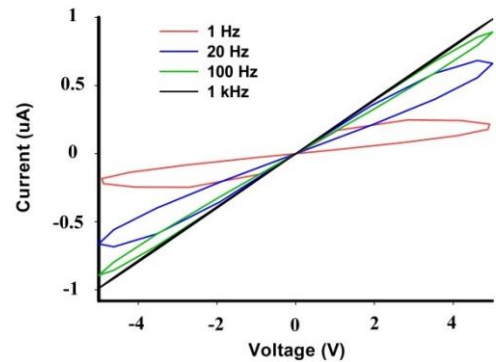

Fig. 1. Pinched hysteresis loop of the memristor

## III. THE BASIC CURRENT MIRROR CIRCUIT

The basic current mirror circuit is presented on the Fig. 2 (a). It consists of two resistors, $R_1$ and $R_2$, and two NMOS transistors, $T_1$ and $T_2$, which is assumed to have exactly same parameters. The $R_1$ and $R_2$ is set to same value. This makes the drain $T_2$ have the same potential as the drain of $T_1$. It also should be noted that $V_{GS1}= V_{DS1}= V_{GS2}$ [5].

For investigation purposes the resistors in the Fig.2 (a) were replaced with the memristors. The result is shown on Fig. 2 (b).

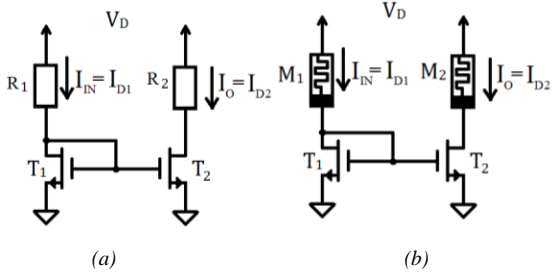

*(a)*        *(b)*

Fig. 2. Basic current mirror circuit with (a) semiconductor resistors; (b) memristors

The behaviour of the circuits presented on the Fig. 2 could be analyzed in terms of relative change in current and its dependence on the relative change in resistance. Two basic equations (3) and (4) are firstly introduced to show the expression of the drain to source voltages in two branches of circuit (a), which can be used to derive the difference between two currents $I_{D1}$ and $I_{D2}$. This difference could be observed in Eqn. (5)

$$V_{DS1} = V_{DD} - I_{D1}R_1 \quad (3)$$
$$V_{DS2} = V_{DD} - I_{D2}R_2 \quad (4)$$

$$\Delta I_D = I_{D2} - I_{D1} = \left(\frac{V_{DD}-V_{DS2}}{R_2}\right) - \left(\frac{V_{DD}-V_{DS1}}{R_1}\right) \quad (5)$$

The relative change in current, which is the difference between $I_{D1}$ and $I_{D2}$ divided by the input current $I_{D1}$, is represented in Eqn. (6).

$$\frac{\Delta I}{I_{D1}} = \frac{\left(\frac{V_{DD}-V_{DS2}}{R_2}\right)-\left(\frac{V_{DD}-V_{DS1}}{R_1}\right)}{\left(\frac{V_{DD}-V_{DS1}}{R_1}\right)} \quad (6.1)$$

$$\frac{\Delta I}{I_{D1}} = \frac{V_{DD}\left(\frac{1}{R_2}-\frac{1}{R_1}\right)+\frac{V_{DS1}}{R_1}-\frac{V_{DS2}}{R_2}}{\frac{V_{DD}-V_{DS1}}{R_1}} \quad (6.2)$$

Assuming that $\frac{V_{DS1}}{R_1}$ and $\frac{V_{DS2}}{R_2}$ are small because $R_1$ and $R_2$ are in kilovolts, the relative change in current could be simplified to Eqn. (7).

$$\frac{\Delta I}{I_{D1}} = \frac{V_{DD}\left(\frac{R_1}{R_2}-1\right)}{V_{DD}-V_{DS1}} = \left(\frac{R_1}{R_2}-1\right)\left(1-\frac{V_{DS1}}{V_{DD}}\right)^{-1} \quad (7)$$

As $R_1$ is taken as a constant resistor value of 38kΩ, $V_{DS1}$ remains constant. This implies that constant K could be introduced instead of $\left(1-\frac{V_{DS1}}{V_{DD}}\right)^{-1}$.

Considering all above mentioned, the relative change in current can be expressed with respect to $\frac{|\Delta R|}{R_1}$, which is relative change in resistance (Eqn. (8)).

$$\frac{\Delta I}{I_{D1}} = K\left(\frac{R_1}{R_2}-1\right) = K\left(\frac{R_1-R_2}{R_2}\right) = K\frac{|\Delta R|}{R_2} = \frac{C}{R_2}\frac{|\Delta R|}{R_1} \quad (8)$$

The same equation can be introduced for memristor by interchanging R with M (Eqn. (9)).

$$\frac{\Delta I}{I_{D1}} = \frac{C}{M_2}\frac{|\Delta M|}{M_1} \quad (9)$$

## IV. CIRCUIT SIMULATION RESULTS

The TSMC 0.18u BSIM spice models and memristor models were used our study [6]. The conventional current mirror circuit with resistors at 27°C with a power supply of 2.5V. The resistor value that was taken is 38kΩ to ensure that the transistor is driven in saturation. The simulation results showed that resulting output and input current match. It can be noted that output current duplicates input completely and this is the ideal case which could be hardly achieved in real life due to manufacturing errors.

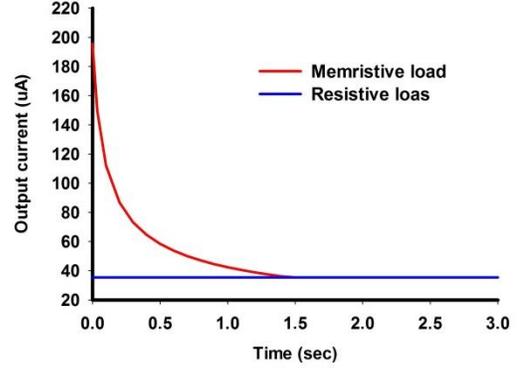

Fig. 3. Simulation results for circuit with resistors and memristors

To study the impact of replacing the resistor loads with memristor, we modify the circuit in Fig. 1(a) with that of Fig. 1(b). The resulting output currents for resistor and memristor is shown in Fig. 3. In contrast to resistors, the switching behavior of the memristors introduces a high current which gradually settles to a constant value. In the Verilog memristor model that was used for the simulation the initial resistance value of the memristor is 5kΩ. When we apply the voltage of 2.5V, the resistance value of the memristor increases up to 38kΩ. This switching period occurs during the first 1.4s for 2.5V input, after this the current value stabilizes and remains unchanged. Like a normal current mirror, the input and output currents for the circuit with memristors are the same.

The switching period of the memristor is an interesting phenomenon as emphasized in the Fig. 4. On this graph the output current for different input voltage values is shown. The initial memristance value for all cases is 5kΩ and changes up to about 38kΩ for all cases. Even though the initial and final memristance values are approximately the same, the switching time for these cases is different and depends on the supply input voltage value. The switching period increases with the decrease of input voltage. This can lead to limitations in future application.

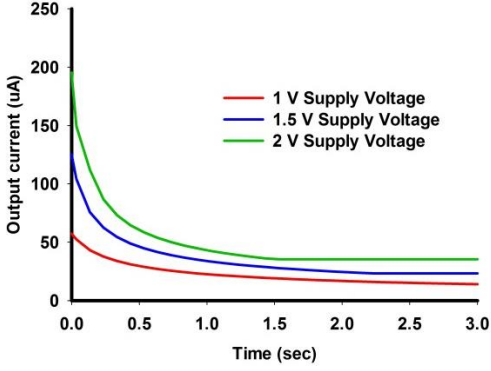

Fig. 4. Output current through the memristor for different input voltages

The percentage change in output current with the change in output resistance for both circuits is shown on Fig. 5. This analysis was done by varying output resistance R2 and keeping input resistance constant. The horizontal axis of the graphs represents the percentage difference between input and output currents, while the vertical axis shows the percentage difference between input and output currents. The change in current for both circuits is the same.

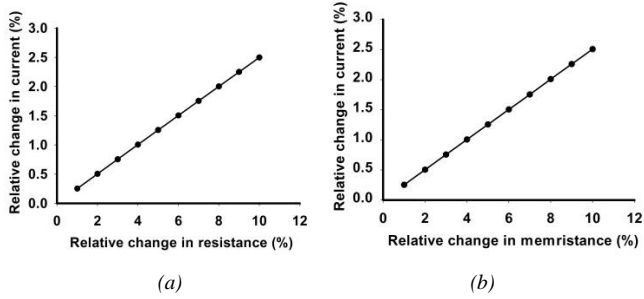

Fig. 5. Relative change in current corresponding to the relative change in resistance for the circuit with (a) semiconductor resistors; (b) memristors

## V. ANALYSIS OF BASIC CHARACTERISTICS

The semiconductor resistors are impacted more to changes in temperature than the memristive current mirrors as shown in Fig 6. The output current is the same as input current for both circuits for all temperatures. In both cases, the current decreases with the increase of temperature. The currents for the circuits with memristors remain far more stable than the semiconductor resistive mirrors.

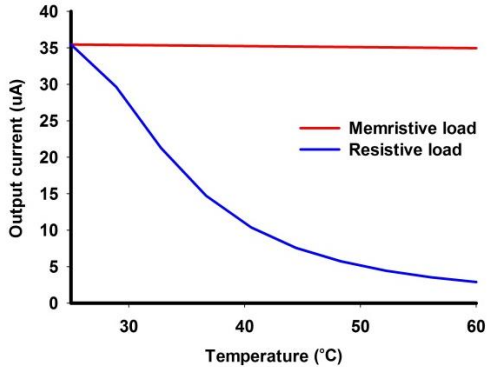

Fig. 6. Effect of temperature change on output current for the circuit with semiconductor resistive and memristive loads

Another characteristic that helps to benchmark the current mirror is the Total harmonic distortion (THD), and it defines the distortion level and the mismatch between input and output parameters [7]. In order to find the distortion level for the current mirror circuit, we changed the supply voltage AC type and performed a Fourier analysis. The input signal is sinusoidal with amplitude of 2.5V and frequency of 50Hz. The results for both circuits are shown in the Table 1. The total harmonic distortion level for the circuit with resistors is slightly higher than for one with memristors. However, this difference is insignificant and does not have a noticeable impact on the results.

Total harmonic distortion (THD) can be calculated by Eqn. 10 which shows THD related to the fundamental (first order voltage) voltage.

$$THD = \frac{\sqrt{\sum V_n^2}}{V_1} \qquad (10)$$

In Eqn. X, $V_n$ corresponds to the harmonic voltage, where n is a harmonics order. $V_1$ is a voltage of the first harmonic [8].

Power dissipation would remain same under the steady state in both the circuits. The power dissipation with a power supply of 2.5V and resistance value of 38kΩ can be seen from Table 1. This is due to the same supply voltage and similar input and output current values. However, in real circuits the power dissipation in the current mirror with resistors is larger due to the resistor leakage current which does not exist in memristors. When calculating the power dissipation in the circuit the leakage power due to the leakage current in the transistors should be accounted for as well.

The subthreshold leakage current for the transistor model cold be expressed in terms of Eqn. (11):

$$I_{sub} = I_0 e^{\frac{V_{gs}-V_{th}}{nV_T}} \left[1 - e^{\frac{-V_{ds}}{V_T}}\right] \qquad (11)$$

where $I_0 = \frac{W\mu_0 C_{ox} V_T^2 e^{1.8}}{L}$ and thermal voltage is expressed as

$$V_T = \frac{KT}{q} \qquad (12)$$

The equation for the gate leakage current is shown in Eqn. (13)

$$I_{gate} = WLA\left(\frac{V_{ox}^2}{t_{ox}}\right) exp\left[\frac{-B\left(1-\left(1-\frac{V_{ox}}{\emptyset_{ox}}\right)^{3/2}\right)}{\frac{V_{ox}}{t_{ox}}}\right] \qquad (13)$$

A and B in Eqn. (13) are the parameters that could be represented by the Eqns. (14) and (15).

$$A = q^3/16\pi^2 h \qquad (14)$$

$$B = 4\pi\sqrt{2m_{ox}}\emptyset_{ox}^{3/2}/3hq \qquad (15)$$

where $t_{ox}$ is the oxide thickness, $m_{ox}$ is the effective mass, $\emptyset_{ox}$ is the tunneling barrier height, $h$ is the reduced Plank's constant, $q$ is the electron charge [9].

Lower area on chip is one of the main advantages of the memristor as shown in Table 1. There is a significant difference between the areas of the circuits with resistors and with memristors. We assume that the used semiconductor resistors have width of 2µm and length of 10µm [10]. The dimensions of memristors were taken as 45nm*90nm considering the maximum on-chip area. This assumption was made according to the existing memristor models [11]. The used transistor model has a length and width of 0.18 µm and 0.27µm.

TABLE I.  CHARACTERISTICS OF TWO CURRENT MIRROR CIRCUITS

| Current mirror configurations with | THD (%) | Power dissipation (mW) | Area (pm$^2$) |
|---|---|---|---|
| 2 resistors | 1.779 | 0.1418 | 17.44 |
| 2 memristors | 1.774 | 0.1416 | 1.5372 |
| 1 PMOS transistor and 1resistor | 1.749 | 0.2048 | 10.88 |
| 1PMOS transistor and 1 memristor | 1.705 | 0.2040 | 2.9286 |

After the analysis of the two current mirror circuit configurations, we decided to implement the circuits with the PMOS transistor used instead of the first resistor and memristor, $R_1$ and $M_1$, in the Fig. 2 and account for the capacitive effects. The results of the modifications in the circuits are shown on the Fig. 7.

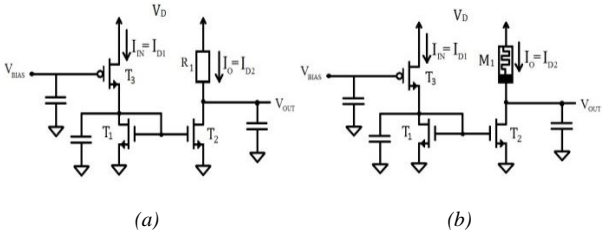

*(a)* *(b)*

Fig. 7. Current mirror circuit configurations for illustration of the capacitive effects of the (a) resistive and (b) memristive circuit

Same types of analysis were performed as before. From Fig. 8 it could be seen that for the 0.7 V bias and 2 V supplied voltage almost the same amount of time is required for the memristor to overcome the switching period. The stabilized current values of memristor based circuit and constant value of the current in the resistive circuit are slightly less than 40uA in this case. The analysis is based on the 0.18 micron transistor technology.

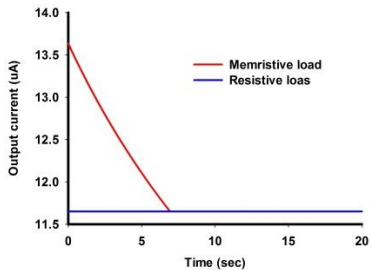

Fig. 8. Output current change with time

As a result of changes in the current mirror configuration, the output current now depends not only on the amount of the voltage supplied, but also on the bias voltage. Fig. 9 and Fig. 10 show the effect of bias voltage change on the output current. It could be seen that with supply voltage kept at constant 2 V, the current value after the transition period gets lower. Moreover, there is clear dependence of the transition period and the difference between the supplied and bias voltage.

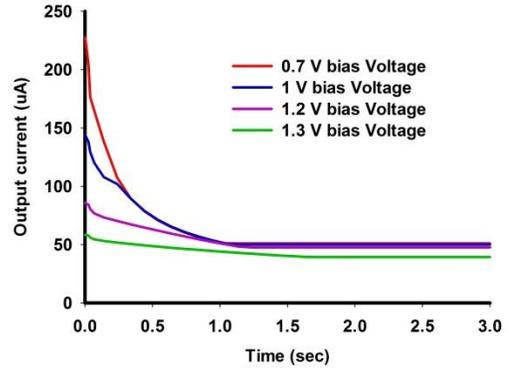

Fig. 9. Change in the output currents with at different bias voltages

The change of output current with bias voltage for the circuit with PMOS transistor and resistor and PMOS transistor and memristor is shown in the Fig.10.

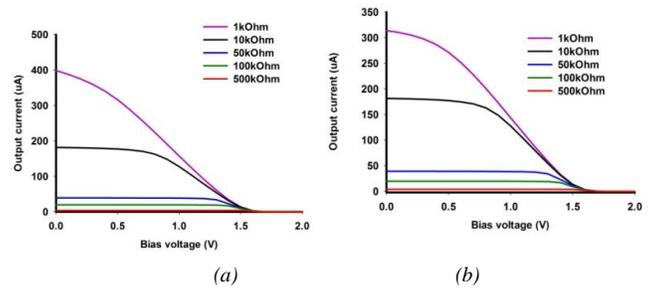

*(a)* *(b)*

Fig.10. Dependence of the output current on the bias voltage for (a) resistor and (b) memristor based circuits

The output voltage of the circuit is also affected by the change in the bias voltage. This is presented in the Fig. 11

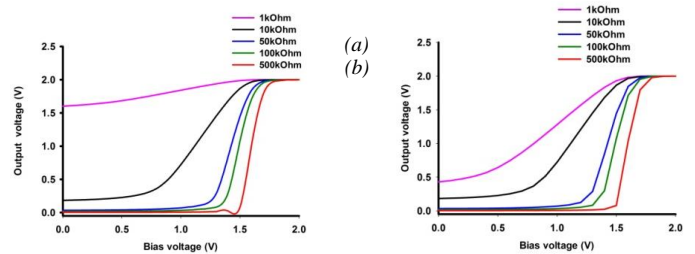

Fig. 11. Dependence of the output current on the bias voltage for (a) resistor and (b) memristor based circuits

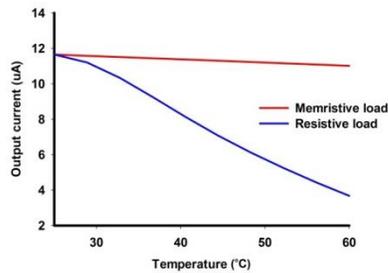

Fig. 12. Effect of temperature change on output current for the circuit with semiconductor resistive and memristive loads

The change in current corresponding to the change of the width of output transistor T2 is shown in the Fig.13.

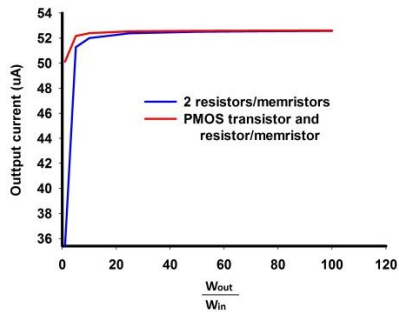

Fig. 13. The change in output current with the width of output transistor T2

The relative change in output current with relative change of threshold voltage of output transistor T2 is shown in the Fig.14.

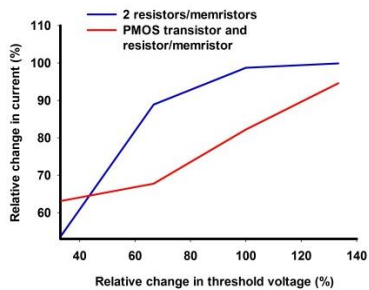

Fig. 14. The change in output current with the threshold voltage of output transistor T2

## CONCLUSION

In this paper, we presented the analysis of the basic current mirror circuit with memristive load. The improved performance of the proposed circuit with that of the semiconductor resistors was reported. We demonstrate the feasibility of using memristors in current mirror circuit with the major advantage of low on-chip area, and lower leakage currents. Our results suggest that the replacement of the semiconductor resistor with the memristor model will reduce the area required for fabrication of the circuit by more than the half. In addition, proposed circuit design showed slight improvement in the THD, while the power dissipation parameter was left unchanged. The future research can include a detailed study on noise performance, parasitic extraction and perform the impact of memristor switching periods.